\documentclass[preprint,12pt]{elsarticle}

\newcommand{\ket}[1]{\left| #1 \right>} 
\newcommand{\gv}[1]{\ensuremath{\mbox{\boldmath$ #1 $}}}

\usepackage{amsmath}
\usepackage{amsfonts}

\journal{Studies in the History and Philosophy of Science}

\begin{document}
\begin{frontmatter}
\title{{\normalsize \textbf{Causation, Information, and Physics}}}
\author{{\normalsize \textbf{Geoff Beck}}}
\address{University of the Witwatersrand, School of Physics,Private Bag 3, WITS-2050, Johannesburg, South Africa}
\ead{geoffrey.beck@wits.ac.za}
\date{}


\begin{abstract}
This work outlines the novel application of the empirical analysis of causation, presented by Kutach,
to the study of information theory and its role in physics. The central thesis of this paper
is that causation and information are identical functional tools for distinguishing controllable correlations,
and that this leads to a consistent view, not only of information theory, but also of statistical physics and quantum
information. This approach comes without the metaphysical baggage of declaring information a fundamental ingredient in physical reality
and exorcises many of the otherwise puzzling problems that arise from this view-point, particularly obviating the problem of `excess baggage' in quantum mechanics. 
This solution is achieved via a separation between information carrying causal correlations of a single qubit and the bulk of its state space. 
\end{abstract}

\end{frontmatter}
\section{Introduction}

Fundamental or provisionally fundamental physical laws have no causal character. This curiosity, prominently remarked upon by Russell~\cite{russell1912}, 
necessitates some elaboration: a notion of causation requires both that there is an asymmetry in determination 
(if $A$ causes $B$ then $B$ cannot be said to cause $A$) and that causes 
are well-defined local factors~\cite{russell1912,kutach2013}. 
Physical law cannot be said to be causal as it does not satisfy either requirement: initial/boundary conditions specified at a given time $t$ can 
be said to determine the infinite past of the system's state as much as they determine its infinite future. Moreover, causal sequences can be falsified without the falsification of any physical law~\cite{russell1912}, thus divorcing any notion of causation from laws themselves. What we are safe in declaring is that, using just their mathematical structure, all ostensibly fundamental physical laws deal
in \textit{correlations}. That is, physical laws answer the question: `if we take some event $C$, what further events (or past events) $E$ could we expect with what probabilities?'. The inclusion of probability here is for the sake of generality, we need not insist on whether or not probabilities could be universally applicable. The sequential associations of events or the evolution of the physical state of some system are then the correlations referred to.
Despite this, notions of \textit{causal correlation} are deeply useful from a practical standpoint, as their importance in the work of many scientific fields,
as well as common sense, attests to. The primitive notion of causation is that two events are more explicitly linked than merely occurring often together. The `cause' event in fact directly induces the `effect' event (or at least makes it significantly more likely to occur than other outcomes), in a manner we can see is 
quite opposed to the ambiguity of fundamental physical laws, which uniformly and symmetrically relate a very large set of events indeed.

In order to reconcile these differences we can turn to the empirical analysis theory of causality due to 
Kutach~\cite{kutach2013}, and take causality to be how we, as scientists and humans, distinguish controllable correlations from those that we cannot control. 
Thus, `cause-effect' relations are excised from our metaphysics, but can still be described in terms of physical laws.
This statement of causality is a highly simplified one and it is worth spending some time in explanation. However, for a truly thorough treatment 
the reader is invited to consult the aforementioned work of Kutach.

This empirical analysis of causation is premised on two core principles: that there is a distinction between some fundamental reality and a derivative level, there is also a distinction between metaphysical and non-metaphysical aspects of causation. This results in a division of reality into three layers for the purpose of this analysis: the first being fundamental reality that is only concerned with the metaphysics of causation, the other two being derivative realities relating each to metaphysical and non-metaphysical aspects of causation. We will only be interested in those layers discussing the metaphysics of causation and will simply refer to them as fundamental and derivative as a result. Fundamental Reality (FR) is characterised by four simple points:
\begin{enumerate}
	\item[1.] FR is how things really are
	\item[2.] FR is the real basis for events in derivative reality
	\item[3.] FR is as determinate as reality gets
	\item[4.] FR is consistent.
\end{enumerate}
The reason these distinctions are useful is that the usual `cause-effect' relation between bits of reality, each with certain characteristics, will be relevant within the derivative levels of reality, it will not be part of the fundamental metaphysics, but is ultimately predicated on the laws of fundamental reality. It will be seen in the following that, `cause-effect' relations depend only on a part of the full fundamental situation, this being selected by means of derivative phenomena. Implying that `cause-effect' relations belong to no one layer of reality, so they are safely excised from our metaphysics, obviating many problems that arise when `cause-effect' relations are held to be metaphysical necessities. This is justified on the basis of the following consideration proposed in \cite{kutach2013}: a ferromagnet consisting of spins is moved near a device that detects currents via the deflection of a needle. The `cause-effect' relation will consist of the motion of the magnet causing the deflection of the needle. However, the movement of the spin system is only one event in a very large set of fundamental events that result in the movement of the needle. We single out the movement event because we believe it more significant (in a counterfactual sense) to the resulting effect than the other related fundamental events. 

To illustrate the distinction of fundamental and derivative realities we can follow \cite{kutach2013} in considering the framework of classical physics (as this was once thought fundamental). This consists of a limited set of fundamental objects needed to define the laws: particles, properties of mass and charge, and a spacetime equipped with a distance relation. These ingredients are a minimal set of all that is needed to explain fundamental classical reality. Other objects that exist or occur as a result of these fundamental existents are derivative as they are neither part of the framing of the fundamental laws nor assumed in them~\cite{kutach2013}. An example of this is velocity in classical physics, as this corresponds to no fundamental structure, needing a frame of reference (not part of the fundamental set of existents) to have any definite value. 

There are, of course, a wide variety of both one~\cite{mackie1973,lewis1973,suppes1970,salmon1977,kistler1999,dowe2000} and two~\cite{good1961,good1962,sober1985,eells1991,salmon1993,hall2004} core concept analyses of causation where `cause-effect' is taken to be a metaphysical necessity. However, the difference here that results in causation being merely an empirically useful derivative phenomenon is that no one of the three layers of reality hosts the whole of a core concept of causation or entirely encapsulates a `cause-effect' relation. In this way, causation can be described in terms of `effective strategies', that is, a causal relation is one that some agent could in principle use to bring about a desired effect. Note that this does not tie causation to agency, as this is merely used to analyse what is permitted by physical law and no agent/agency is necessary to the existence of causal relations. The agency can be better understood as the freedom allowed by physical laws for any process to realise some outcome in principle.  

This approach to the study of causation will be formalised in this work, and the mathematical formalism will then be leveraged to address several key problems in 
the overlap of physics and information theory. This will be done by demonstrating that the notion of information used in physics is identical to the definition of causation in terms of empirical analysis. Once this identification has been made it becomes possible to obviate the circularity in Shannon's original definition of distinct states~\cite{shannon1948} and information. This is because the notion of what states can carry information is now formulated in terms of available causal correlations, which depend only on the possible counterfactuals as well as the prescriptions of fundamental physical laws in a manner similar to that recommended in \cite{maudlin2007}. 

The identity between causation and information is then used to address the problem of `excess baggage' in quantum mechanics~\cite{hardy2004}, where states (if they are in some way ontic) must seemingly carry both finite and infinite amounts of classical information~\cite{leifer2014,leifer2015}, in contravention of the established finite limits of the Holevo theorem~\cite{holevo1973}. The correspondence of information and causation is used to tease out this problem and demonstrate that only the finite information carried by the quantum state actually corresponds to causal correlations of a single qubit, the infinite component is accommodated by an ensemble of qubits only. This is of particular importance to the discussion of the ontic status of the quantum wavefunction, as it would otherwise seem impossible for an ontic wavefunction, as discussed in \cite{psi1,psi2,psi3,psi4,psi5,psi6,psi7,timpson2010}, to contain infinite information and still obey the Holevo theorem. Here the full utility of Kutach's account of causation is realised, in that this solution is only available when information can be regarded as an operational tool only and not being part of our metaphysics. As is shown here, this can be achieved through the identification of causation and information that is possible through Kutach's empirical account.

This paper is structured as follows: In section~\ref{sec:causes} the causal formalism is layed out and discussed. Section~\ref{sec:info} outlines the link between causation and information in physics, and section~\ref{sec:baggage} addresses the solution of excess baggage problem. 


\section{Causation}
\label{sec:causes}

The first ingredient in this analysis of causation is the distinction between fundamental and derivative reality. 
The latter of the two refers to the realm of experience
similar to the `classical world' of common physical terminology. Whereas fundamental reality is the domain of the basic forms of 
physical law, quantum theory or any provisionally fundamental theory for instance. The distinction is simply that the laws of derivative reality
must be said to be determined by those of the more fundamental underlying reality.
This distinction informs the way we talk about events within these two realities, those that occur in derivative reality being the only cases
where we must explain why causality might be useful or apparent. To this end we classify the events of derivative reality as \textit{course-grained}, 
such an event $E$ is a collection of fundamental events that might correspond to the derivative event, while a \textit{contextualised event} 
is course-grained with a reasonable probability distribution
over its members. In Kutach's presentation, this distribution need not be empirical
or rigorous in any way (in order to accommodate informal notions of causality), 
barring that it must satisfy the axioms of probability theory. 

In order to determine if two time-ordered events, $A$ and $B$ are causally 
correlated we will employ the following terminology. 
The \textit{protrast} of the ordered pair ($A$,$B$) is a set of fundamental events in an event $A$ that would correlate with $B$ being said to occur, whereas the 
\textit{contrast} of ($A$,$B$) is a set of events in a reasonably chosen contrasting 
event $C$ that correlate with $B$ occurring, equivalent to imagining the causal pair ($C$,$B$) instead. We can also, of course, consider causation in terms of a derivative event $A$ and the probability it evolves into $B$ under apparent laws of derivative reality. 
At this point it is worth remarking upon the fact that this formalism illustrates the very close linkage of causality and counterfactuals.
We note that Lewis~\cite{lewis1973} championed analysis in terms of counterfactuals as giving a complete account of causality, but the literature is littered with 
difficulties that this program encounters (as described in \cite{maudlin2007,butterfield1992} for example). We will see, in the course of this paper, that the empirical account of causation employed here
has far greater similarity to the approach advocated by Maudlin~\cite{maudlin2007}. Whereby, physical law provides the connection between
counterfactuals and causation, and is thus the vital ingredient needed to give a complete account. Thus, the notion of a `reasonable choice'
of the contrasting event $C$ is necessarily bound up in what counterfactuals would be allowed by physical law. Since this work is focussed upon causation within formal systems
of physics and information, all counterfactual choices will not just be predicated on physical laws, but rather admissible counterfactuals 
are in fact given directly by physical laws of the system in question. In this sense, the framework for causation here is more strict than it's parent in Kutach's work. However, this is just
as a result of a narrowing of focus for application to particular issues, rather than the general explication of the metaphysics of causation that is the object of the empirical framework itself. This
will prove important as this restriction is vital to be able to define some notion of `counting' of causal correlations within a given physical system.

In the context of an ordered pair of contextualised events $C$ and $E$, the probability of 
$E$ given $C$ is written as $P(E|C)$.
The nature of these probabilities will be taken as the proportion of fundamental events in $C$ which will evolve into those within $E$, meaning there is a robust link between these probabilities and objective frequency measurements. We can naturally expect a fully general discussion to be couched in terms of probabilities as not all of the fundamental events in our course grained event $C$ will necessarily correlate with members of the same event $E$. The use of probability is then motivated to account for the fact that an event $C$ may result in several different possible consequent events and the frequency of differing outcomes will depend upon the strength of the underlying fundamental state correlations. 
Kutach then invokes the notion of \textit{promotion}: the degree to which $C_1$ promotes $E$ is given by the difference between the propensity with which events in $C_1$ and those in a contrasting event $C_2$ would evolve into those in $E$. This propensity can be determined either through fundamental or derivative laws (as these should agree on matters of empirical outcome promotion).
We then propose to formalise the degree to which $C_1$ promotes $E$ as the logarithmic difference (we note that \cite{kutach2013} employs a linear difference)
\begin{equation}
\mathcal{C}(C_1,E) = \log\left(\frac{P(E|C_1)}{P(E|C_2)}\right) \; .
\end{equation} 
This can be generalised to a larger set of events that might cause $E$, rather than simply the two in $\{C_1,C_2\}$. 
To do so consider a set $\{ C_i \; \vert \; i \in [1,N]\}$, then 
\begin{equation}
\mathcal{C}(C_n,E) = \sum\limits_{i \ne n}^{N} w_i\log\left(\frac{P(E|C_n)}{P(E|C_i)}\right) \; , \label{eq:many}
\end{equation}
where $w_i$ is a weight, given by $\frac{1}{N}$ if we cannot define the probability $P(C_i|\mathcal{L}_{phys})$, where this is conditioned on the relevant set of physical laws $\mathcal{L}_{phys}$ or if we consider a
scenario such as a symbols transmitted freely on some communication system.
The function $\mathcal{C}(E,C)$ is then proposed as a measure of the causal association of the 
events $C$ and $E$. Under this assumption, if $\mathcal{C}(C,E)$ is positive-definite we may conclude that it is reasonable to
state that $E$ is causally correlated with $C$. However, if $\mathcal{C}(C,E) \le 0$  
then we must conclude that on average other events that promote $E$
equally or to a greater degree than $C$ does, making causal claims about $(C,E)$ weak.  
The relative magnitude of the $\mathcal{C}$ function will also dictate the extent of the causal association between the two events.
This causal association can be understood as follows: two events $C$ and $E$ are causally associated if there exists a physical scenario
whereby the event $C$ occurring would offer a preponderant probability (over and above most other strategies) of $E$ being
a consequence of $C$.

There are two approaches to continuous families of events, the first is a simple generalisation of Eq.~(\ref{eq:many}),
\begin{equation}
\begin{aligned}
\mathcal{C}(E,C) & = \int dc \; \log\left(\frac{P(E|C)}{P(E|c)}\right) \; . \label{eq:cont}
\end{aligned}
\end{equation}
However, this presents a difficulty in enumerating causal correlations.
Therefore, the continuum should be reduced to a discrete case by identifying causal classes of events.
In general an event $A$ in a continuous family can be characterised by 
some set of parameters $\gv{\eta}$, it seems appropriate to determine whether $A(\gv{\eta_1})$ and $A(\gv{\eta_2})$
are contrasting events by the `outcome continuity' of $\gv{\eta}$. Thus if one can continuously deform $\gv{\eta_1}$ to obtain $\gv{\eta_2}$, without causing a change
in the most probable outcome of $A(\gv{\eta})$, then the two events
must be seen to belong to the same `contrast class', meaning they cannot be chosen as contrasting events because they are causally equivalent. 
For a larger set of outcomes a contrast class with constitute all events $A(\gv{\eta})$ that preserve the same probability heirachy.
Continuity is important because the value of $\gv{\eta}$ will serve to demonstrate control
of a particular correlation, you can `tune' $\gv{\eta}$ to more strongly promote a given outcome while conducting what 
is ostensibly the same experiment.
We can appreciate the use of `outcome continuity' if we view a lack of this as signalling 
that these events assign the greatest probability to differing causal histories, making $\gv{\eta_1}$ and $\gv{\eta_2}$ causally distinct.
We note that $\gv{\eta_1}$ and $\gv{\eta_2}$ might anyway result in the same event history but this does not
damage the use of `outcome continuity' as the argument is probabilistic in nature.

In order to take the concept of the contrast class into account, any probability $P(E|A)$,
where $A \equiv A(\gv{\eta})$, will be assumed
to be averaged over the relevant continuous region of $\gv{\eta}$ unless otherwise stated. 
The reason for this is to make causal arguments robust and not simply dependent on the choice of parameters, which is of particular concern in
the contrasting causes, as these might otherwise be chosen to minimise their association with a given outcome.
This attempt to characterise contrasting events is aimed at allowing physical law and operational considerations to determine our
contrasting event classes, in keeping with the important role of physical law in determining the counterfactuals necessary in causation.
It is very clear that if we are to make an empirical analysis of causation that we should only admit causal counterfactuals 
that would stand up to empirical inspection. 


In order to make full use of the function $\mathcal{C}$ we can make use of a causal table. Illustrated below for a system with two events $C_1$ and $C_2$ with two possible outcomes $E_1$ and $E_2$.
\begin{center}
\begin{tabular}{l|ll}
 & $E_1$ & $E_2$ \\
\hline
$C_1$ & $\mathcal{C}(E_1|C_1)$ & $\mathcal{C}(E_2|C_1)$ \\
$C_2$ & $\mathcal{C}(E_1|C_2)$ & $\mathcal{C}(E_2|C_2)$ 
\end{tabular}
\end{center}
In this table we can scan down column $i$ to pick out possible causes for event $E_i$. These can then be tested by scanning across the row of a favoured cause to see that it does
not uniformally promote multiple outcomes.

The method of enumerating causal correlations in a given system requires remarking upon. For a given correlation we might always pick the largest $\mathcal{C}$ value and decide it is the only causal correlation. However, this cannot be correct, as a simple example can show. Consider a configuration of $N$ molecules that results in a total energy $E$, which correlates with some additional observable values. In principle there are many such configurations, each of which has probability $1$ of associating itself with measurements of the observables that correlate to $E$. Thus, among this set of correlations all have $\mathcal{C} = 0$ as we cannot prefer any of them. Moreover, we might consider all these states with energy $E$ as one single causal class. It is evident that this difficulty arises due to the determinism of the problem, in that states have either $P = 1$ or $P = 0$. Additionally, one can appreciate that $P = 0$ states cannot be considered valid counterfactuals, so they cannot be included to make $\mathcal{C}$ non-zero for those with $P = 1$. However, this difficulty can be resolved simply because of the determinism, all of the $P = 1$ correlations are causal, as though they have $\mathcal{C} = 0$, they are deterministic and there are no other valid contrasts to consider. Thus, the process of enumeration must be cautious, for a given effect we will take the causal classes with positive $\mathcal{C}$ values to be the set of causal correlations that produce this effect, in the case that all have the same $\mathcal{C}$, or there appear to be no causal correlations at all, we must carefully inspect the $P$ values to confirm any conclusions.

Thus, we aim to present the causality measure $\mathcal{C}$ 
as a formal and mathematical realisation of Kutach's promotion causality, allowing it to be used in more specialised physical discussion as well as
assessments of the general use of causality. 

It is evident that Kutach's theory gives rise 
to the notion of causality as a functional tool used to make the distinction between correlations we can control and those which we cannot. Thus, we can 
see that causality can be apparent within derivative reality without being a necessary ingredient of fundamental reality.
What must also be clear is that our causal expectations will be recovered only if
we select counterfactuals that are allowed by physical law and assign them physically sensible weightings, 
arbitrary fantasy counterfactuals could, for obvious reasons, easily undo our reasonable causal expectations (as argued by Maudlin~\cite{maudlin2007}).

\section{Information}
\label{sec:info}
To study the notion of information we will approach it from the perspective of communication and how this relates to the underlying physics. 
To do so we must define some terms, the first being a \textit{dictionary}: this is a set of symbols which are
assigned to some states within a physical system, each symbol having some meaning which we are free to choose when specifying the dictionary. 
For example we can encode a binary dictionary onto a current being measured in a wire, no current detected is assigned the symbol
$0$ while detection above a given threshold is assigned $1$. 
If we then measure the current for a length of time we can translate this into a serious of $0$'s 
and $1$'s which may be interpreted as a  which may contain information. 
By determining the number of distinguishable states available to our transmitting system we can calculate the Shannon entropy of a communication channel,
in the binary case we have two available states and thus a message of $N$ symbols on our channel has $S \propto N\log{(2)}$ which provides an information measure 
when the results of measurements on our system are treated as values of a random variable. In this regard, the information ascribed to a message sent on some physical channel 
can be considered as the apriori degree of unpredictability of the constituent symbols in the message.  
To discuss the relation of causality and information let us first enumerate the correlations in our binary system: 
we can have generation of current by some process at one end of the wire that correlates with detection of a similar amplitude current at the other end, 
or the case of no current being generated being correlated with below-threshold current detected at the other end. 
However, there are other correlations available to the binary system: these being the cases of `mis-correlation', where we detect a current without 
being correlated to generation or where we detect no current despite a current being generated.
At least some of these `mis-correlations' would be physically justifiable, so we must ask under what conditions can
we use the distinguishable states of a system to encode information. 
It is then clear that if we wish to use the distinguishable states of a system to transmit information then we must be able to reliably induce
particular correlations in that system. Otherwise, the very notion of the `distinguishability' of these states is lost.  
In the binary system for instance, if we cannot reliably induce a current 
that is found upon subsequent measurement to be above the given threshold then we are in
danger of scrambling any message on our channel, as our $1$'s might frequently appear as $0$'s (and thus no longer be clearly distinguished between). 
A particular physical system $X$ can then be said to be capable of
transmitting/containing information if we can reliably map some dictionary onto a subset of possible correlations within that system.  
We contend that this subset is composed of only the causal correlations of the system in question. 
This can be demonstrated using the terminology established in the previous 
discussion. Let $A$ be the process we use to attempt to induce a particular physical state $x$ in the system $X$ and $B$ be the realisation/measurement 
of that state. Then, if we have a case where $\mathcal{C}(A,B) \leq 0$, or $\vert \mathcal{C} \vert \leq \epsilon$, we must conclude that whether or not we can 
realise our desired state $x$ through the process $A$ is highly mis-correlated and cannot be said to produce a distinguishable state. 
Therefore, if we attempt to transmit/store a sequence of symbols with $X$ it will be akin to a stream of bits where 1 and 0 are frequently interchanged, garbling any message we might send and thus preventing us from transmitting any desired information.
Clearly if all correlations available to the system have $\mathcal{C} = 0$ (and are confirmed non-causal) we must conclude that there can be no information 
transmitted/stored. Every added correlation class with $\mathcal{C}$ significantly greater than zero must therefore expand the possible information content of messages
realised within the system, as we can reliably expand our dictionary with each added causal correlation.
We can therefore conclude that for a correlation to contain information it must necessarily be a causal correlation and conversely that any causal correlation 
may store or transmit information. Thus we propose that the information capacity of a physical channel is framed in terms of the channel's causal correlations rather than distinguishable states.
Importantly this definition of information clears up the problem of circularity in Shannon's original definition:  
Information can only be encoded in/transmitted via causal correlations of some physical system and these are defined by the measure $\mathcal{C}$ and thus by the possible counterfactuals
and probabilities derived from our theory of physical law. This means that, although causal correlations are those that can carry information, they are not
defined as such and thus any circularity is obviated. Additionally, the definition of causal classes provides a natural way to
obviate problems of distinguishability in continuous variable systems, by differentiating between them via their promotion of outcomes.  
Furthermore, we also find that the Shannon entropy can be determined through the counting of causal correlations available to the system.
This is because this counting is degenerate with that of distinguishable physical states when these are members of derivative reality 
(this being the domain of `classical' information theory).
However, the examples presented in the remainder of this work will demonstrate that the counting of causal correlations provides a far more robust and 
consistent measure of information content.
Generalising the counting of causal correlations, by analogy with Shannon theory~\cite{shannon1948}, the entropy becomes
\begin{equation}
S = -\sum\limits_{i=1}^{N} p_i \log \left(p_i\right) \; ,
\label{eq:entropy}
\end{equation}
where the sum runs over the causal correlation classes corresponding to the contextualised pairs ($A_i$,$B_i$). 
The weight $p_i$ of each correlation class
is the probability of the correlation being realised.

The preceding arguments suggest a striking agreement between the notions of derivative empirical causation and information.
It follows that this equivalence implies that there is nothing `informational' in the laws 
that govern our fundamental reality, just as these laws are not causal. This must follow from the notion that causality is not
so much a property of any fundamental reality as it is a tool for its analysis, so information is not a physical property of correlations as much as it
is a flag of the controllability of said correlations. This stands at odds with a prevailing school of thought within the physical sciences
which champions the `information is physical' view-point, notably articulated by Brillouin~\cite{brillouin1952} and Deutsch~\cite{deutsch2014} among others. 
In this paradigm information is a fundamental ingredient in laws of physics, and that information itself is an essentially physical quantity. 
For this reason we must supply an argument as to why information seems sufficiently significant in physics as to warrant such extraordinary metaphysical
assertions, while being simply a tool of studying reality.
It must be immediately apparent that this is answered by the entire premise of the presented model of causality, or simply put: 
information theory is so applicable in sciences because its very formulation guarantees it to be so. 
In fact the nature of information capacity as a demarcation between useful and non-useful correlations makes it impossible that it would not be applicable 
to the study of correlation and regularity that tends to compose the majority of sciences. We note that this does not justify the scientific status of the use of
causality/information, but merely explains it. The justification of the use of causation in scientific endeavour will be examined in future work.

In this particular work we apply this causal account of information to problems in quantum information theory. This means we must ask if this notion of information
is adequate to the task. We can immediately see that this causal account of information, being structured on the division between fundamental and derivative reality, 
is immediately well suited to quantum mechanical problems which involve extracting information via measurements conducted in derivative (classical) reality from where it is transmitted/stored 
in states that are members of fundamental (quantum) reality.

\section{Excess Baggage and Quantum Information}
\label{sec:baggage}
In quantum mechanics it has been demonstrated that a `qubit' system with two measurable states, 
referred to as `up' or $\ket{\uparrow}$ and `down' or $\ket{\downarrow}$, 
possesses a vast space of possible quantum states~\cite{montina2008}. Commonly this is interpreted as meaning we should
be able to encode a huge amount of retrievable/`classical' information in such a system~\cite{hardy2004,leifer2014} 
as this should scale with the size of the state space (according to the notion that information is a property of distinguishable states). 
This becomes remarkable when it is observed
that we cannot retrieve any more than one bit from such a system, as argued by the Holevo theorem~\cite{holevo1973}. This is immediately problematic if the state of the system is viewed in a realist fashion as the apparently real state space does not correspond to real retrievable information. All the non-retrievable information
is thus referred to as `excess baggage' by Hardy~\cite{hardy2004} and it must be explained why such a vast state space can offer up
so little information. This problem can be viewed as follows: the states of a qubit can be expressed as the points on the surface of a 
sphere of unit radius, known as the Bloch sphere. Thus, they are a function of two continuous parameters. This means that, for a given qubit,
the probabilities of measuring `up' or `down' vary continuously depending on the basis we choose to measure the qubit in~\cite{leifer2015,montina2008}.
This suggests that for every possible basis the qubit represents a different statistical preparation of a classical bit (referred to as a `bit' hereafter).
Therefore, a qubit can be represented by a continuous infinity of bits and must contain an arbitrarily large amount of information. Clearly then there is
some disagreement between how much information a given qubit preparation actually contains, and it is necessary to provide an information measure
that is consistent with both the Holevo bound and the argument outlined above. One resolution to this is to assert that the quantum state has no direct relation
to any underlying ontology. However, in this work we will consider a solution independent of ontic or epistemic assertions. Despite this, a solution 
to the problem of excess baggage is of prime importance to ontic formulations of quantum mechanics, where the quantum state $\psi$ is a part of physical reality
and not merely a summary of information available to some observer (as in an epistemic formulation). It may, of course, be possible to avoid the problem of excess baggage 
by suggesting that although a physical system may occupy a superposition state (in an ontic view of states) it does not immediately entail that the property values of the qubit are in reality
superposed as well. As this would mean that we need not account all the possible property values of the qubit when we construct our information measure and the infinite component of the information thus vanishes. However, the solution we present here is available without need to further justify considerations such as the aforementioned. The value of
our approach is then that it is more strongly general and made available via the empirical causal analysis of Kutach. 

Having established the problem we can mobilise the machinery developed earlier by realising that our coarse-grained events are 
the preparation and subsequent measurement of the qubit. This means that we will look at what causal correlations or information transmission is possible with the set of events: preparation of  some FR state with desired measurement probabilities and measuring one of two qubit values with some classical apparatus. Additionally, our dictionaries are being chosen as follows:
bit value `1' maps to $\ket{\uparrow}$ and `0' maps to $\ket{\downarrow}$. This means that the dictionary is actually chosen when we choose basis, as this
decides what we are measuring when we speak of $\ket{\uparrow}$ and $\ket{\downarrow}$. Thus, rather than being represented by a continuous infinity of bits,
a qubit is in fact compatible with a continuous infinity of seemingly independent dictionaries. It must be remarked upon that the same case cannot be made for a bit,
as even though we could describe a given bit in any basis we please, the commutativity and non-contextuality of it's algebraic description lead all bases to map unambiguously between 
their particular $\ket{\uparrow}$ and $\ket{\downarrow}$ states. The qubit, however, exhibits both non-commutativity of it's observables and the measured values of ``up" and ``down" are strongly basis dependent (if we consider ``up" and ``down" as poles on a sphere, a choice of basis corresponds to a choice of the polar axis). The properties of the operator algebra thus lead to a lack of unique mappings between $\ket{\uparrow}$ states in different bases.
In other words, a measurable $\ket{\uparrow}_1$ state in a given basis could map to a linear combination
of the measurable $\ket{\uparrow}_2$ and $\ket{\downarrow}_2$ states in another basis. This means that measuring $\ket{\uparrow}_2$ cannot be unambiguously mapped to
the result of a counterfactual measurement in the basis defined by $\ket{\uparrow}_1$ and $\ket{\downarrow}_1$. This would seem to suggest that we should be able to 
represent the qubit as an infinite set of independent bits (one per basis). However, the lack of consistent counterfactual statements regarding the choice to measure in different bases will inevitably lead
us to find that causal correlations in one basis will be incompatible with those in other bases and, by implication, the information content of a
qubit is basis-dependent.
Therefore, as a first step we can establish that the reason the qubit appears to be represented by infinite classical bits stems from the non-commutativity of its observable operators and the resulting basis-dependence of what we are measuring when we speak of $\uparrow$ and $\downarrow$ values.

If we examine the causal correlations of a qubit in a given basis then we can see that the Holevo bounds emerges as follows: 
the causal correlations of a qubit in some basis depend on continuous parameters and can be divided into two equally weighted classes, 
`prepare mostly up and measure up' and `prepare mostly down and measure down', these are separated by a discontinuity, in the form of a
class of random correlations that cannot belong to either of the
aforementioned causal classes, as they do not differ in their promotion of contrasting outcomes. This means that the causal correlation space of a qubit in a given basis 
matches that of the statistical preparation of a bit. Therefore, since we must choose a basis to measure in, we will always find that
the qubit can at most yield up one bit of information upon measurement. If we choose to measure in a different basis 
we will find that the causal correlation spaces of different bases are 
not necessarily compatible. For instance, if we define two bases $\beta_1$ and $\beta_2$ such that $\beta_2$ is rotated by an angle $\theta$ along one of the
Bloch sphere directions. We then prepare a qubit so that we can transmit one bit via the causal correlations of $\beta_1$.
The $\ket{\uparrow}_1$ and $\ket{\downarrow}_1$, which are causally associated with our preparation, 
each correspond to superpositions of $\ket{\uparrow}_2$ and $\ket{\downarrow}_2$. 
This means that we find that it becomes highly unreliable to retrieve the $\beta_1$ bit by measuring in $\beta_2$ but also 
that encoding a $\beta_2$ correlation with our $\beta_1$ preparation is just as unreliable (in the sense that we lose distinguishability of states). This is an important issue,
the two bits in bases $1$ and $2$ are not truly independent, so the infinite set of dictionaries do not
in fact encode an infinite set of bits.

This can be fully illustrated by considering a state $\ket{\psi}_1 = \ket{\downarrow}_1$. In basis $2$ there is a probability $\propto \sin{\left(\theta\right)}^2$ of a measurement yielding
$\ket{\uparrow}_2$. Thus, we can see that as we increase $\theta$ we are merely travelling through the causal class `prepare mostly down and measure down'. As we reach some $\theta^{\star}$ we transition into the random class and afterwards proceed into the `prepare mostly up and measure up' class. Thus it is clear that the bases $1$ and $2$ do not posses different causal correlations, they just represent a rotation of the causal class chosen in the preparation basis. Thus, regardless of what basis we choose, we never increase the number of causal classes available to the qubit, there are always just two.

Effectively the basis-dependent behaviour of the qubit observables both seems to add the potential to set up simultaneous `multi-bit' causal correlations as well as providing 
the linkage between bases that prevents any attempt to do so. This makes it clear that we cannot 
independently encode multiple bits upon a single qubit, 
and that the basis-dependence of the qubit observables leads it to appear to be composed of infinite classical bits while still obeying the Holevo bound. 

It is worth noting that we could encode multiple independent bits in the superposition structure of a qubit state in some $\beta_*$, though doing so does nothing to change the number of causal correlations available to the qubit, as our argument above does not depend upon superposition details. However, this preparation does change the causal correlations available to an ensemble of such qubits. This is the case because we cannot extract superposition structure in a single measurement without a Holevo violation. In practice we must perform quantum state tomography and make many measurements to reconstruct the wavefunction. The amplitudes of the states in superposition can then be used to carry information in the same manner as a string of digits. However, these strings are not accessible without an ensemble of qubits to perform tomography upon. In this sense we would not truly increase the information content of a single qubit, as its causal space is unchanged, merely we have exploited the larger causal space of a qubit ensemble. Why is this the case if the information is in $\psi$ which describes the qubit? It is because $\psi$ describes the situation of a qubit with a given preparation in a given measurement process, i.e. it details the entire experimental arrangement and the extra information is being encoded in the statistical relationship between qubits in the ensemble. 

We note that there is a strong similarity between our solution here and that discussed by Timpson~\cite{timpson-qm}. In that Timpson suggests the excess baggage arises from the difference in the amount of information needed to fully specify the state of a quantum system and the amount of information accessible via measurement (this distinction is argued not to arise in the classical case). It is clear that the solution presented here realises a very similar scenario but does so via the use of the empirical analysis of causation to justify the distinction and simultaneously why excess baggage does not arise in a classical context. 

An important aspect of the resolution of the excess baggage problem is that it is completely independent of ontological preference.
Thus it obviates the difficulties experienced by $\psi$-ontology in this regard~\cite{leifer2014}. However, it is worth noting that it does not
then favour any particular ontological/epistemological view-point in quantum mechanics. The empirical/operational nature of the approach makes it
agnostic towards interpretation or metaphysics. 

This mode of explanation seems to break down when we include quantum entanglement, in which case 
we can perform super-dense coding with shared entanglement~\cite{bennett1992} and can retrieve $2 N$ bits from $N$ qubits. However, the shared entangled state
has merely increased the number of exploitable correlations 
and this scenario remains within the remit of the interpretation of information given here. In this scenario we have two sets of qubits, one held by the receiver and one by the transmitter. These two sets are entangled, with the sender and receiver both knowing the nature of the entanglement. The receiver decodes two bits when sent a single qubit because the entangled state basis has four states and four causal correlations. The need for shared entanglement means that
we never expand the causal classes of a single qubit, as the qubit itself does not carry the information of the shared entanglement. This is the case because, without apriori knowledge of the entanglement, we cannot deduce it's presence/nature from a single set of transmitted qubits, we need to compare ensembles to identify the entanglement. In this regard it is similar to the superposition structure case above, the extra information is being carried in the correlations between different sets of qubits. But, in this case, we are also given knowledge of how our qubits correlate with the qubits that are transmitted to us. Thus, our causal state space is built up from two qubits and their correlations. The information content of a single qubit system is unchanged and the addition of extra causal correlations is in keeping with our causal explanation of information transmission. 

What becomes evident is that the causal correlation
view of information applies to all retrievable information, as observable correlations all live within the realm of derivative reality,
in the vocabulary of quantum mechanics they are `classical' objects. This illustrates that the term `quantum information' arises as a result
of the asymmetry between the parameter space of quantum states and that of causal correlations associated with those states. 

It is evident that this view can still be reconciled with those expressed by
Cerf and Adami~\cite{cerf1996}: that quantum correlations, being members of fundamental reality and possessing `quantum information' 
give rise to correlations with `classical information' within derivative reality, although it can only lead to their conclusion that
quantum information gives rise to classical information in the sense that some quantum correlations can map directly onto causal relations.

\section*{Funding}

The author acknowledges support, through a post-doctoral grant, by the South African Research Chairs Initiative of the Department of Science and Technology and National Research Foundation and by the Square Kilometre Array (SKA). 

\section*{Acknowledgements}

I would like to thank Jacques Naude (as well as other `Shannon Day' regulars), Justine Tarrant, Simon Beck, and Douglas Kutach for their input, discussion, and suggestions.

\bibliographystyle{unsrt}
\bibliography{causation}

\begin{thebibliography}{10}

\bibitem{russell1912}
B.~Russell.
\newblock On the notion of cause.
\newblock {\em Proceedings of the Aristotelian Society}, 13:1, 1913.

\bibitem{kutach2013}
D.~Kutach.
\newblock {\em Causation and Its Basis in Fundamental Physics}.
\newblock Oxford University Press, 2013.

\bibitem{mackie1973}
J.~L. Mackie.
\newblock {\em The Cement of the Universe}.
\newblock Oxford University Press, 1973.

\bibitem{lewis1973}
D.~Lewis.
\newblock Causation.
\newblock {\em Journal of Philosophy}, 70:556, 1973.

\bibitem{suppes1970}
P.~Suppes.
\newblock {\em A Probabilistic Theory of Causality}.
\newblock North-Holland, 1970.

\bibitem{salmon1977}
W.~Salmon.
\newblock An 'at-at' theory of causal influence.
\newblock {\em Philosophy of Science}, 44:215, 1977.

\bibitem{kistler1999}
M.~Kistler.
\newblock {\em Causalit\'e et lois de la nature}.
\newblock Vrin, 1999.

\bibitem{dowe2000}
P.~Dowe.
\newblock {\em Physical Causation}.
\newblock Cambridge University Press, 2000.

\bibitem{good1961}
I.~J. Good.
\newblock A causal calculus i.
\newblock {\em The British Journal for the Philosophy of Science}, 11:305,
  1961.

\bibitem{good1962}
I.~J. Good.
\newblock A causal calculus ii.
\newblock {\em The British Journal for the Philosophy of Science}, 12:43, 1962.

\bibitem{sober1985}
E.~Sober.
\newblock Two concepts of cause.
\newblock In P.~Asquith and P.~Kitcher, editors, {\em PSA 1984}, volume~2, page
  405. Philosophy of Science Association, 1985.

\bibitem{eells1991}
E.~Eells.
\newblock {\em Probabilistic Causality}.
\newblock Cambridge University Press, 1991.

\bibitem{salmon1993}
W.~Salmon.
\newblock Causality: Production and propagation.
\newblock In E.~Sosa and M.~Tooley, editors, {\em Causation}. Oxford University
  Press, 1993.

\bibitem{hall2004}
N.~Hall.
\newblock {\em Two Concepts of Causation}.
\newblock MIT Press, 2004.

\bibitem{shannon1948}
C.~E. Shannon.
\newblock A mathematical theory of communication.
\newblock {\em Bell System Technical Journal}, 27:379, 1948.

\bibitem{maudlin2007}
T.~Maudlin.
\newblock {\em The Metaphysics within Physics}.
\newblock Oxford University Press, 2007.

\bibitem{hardy2004}
L.~Hardy.
\newblock Quantum ontological excess baggage.
\newblock {\em Stud. Hist. Phil. Mod. Phys.}, 35(2):267, 2004.

\bibitem{leifer2014}
M.~S. Leifer.
\newblock Is the quantum state real? an extended review of $\psi$-ontology
  theorems.
\newblock {\em Quanta}, 3:67, 2014.

\bibitem{leifer2015}
D.~Jennings and M.~S. Leifer.
\newblock No return to classical reality.
\newblock {\em Contemporary Physics}, 57(1):60--82, 2016.

\bibitem{holevo1973}
A.~S. Holevo.
\newblock Bounds for the quantity of information transmitted by a quantum
  communication channel.
\newblock {\em Problems in Information Transmission}, 9:177, 1973.

\bibitem{psi1}
M.~F. Pusey, J.~Barrett, and T.~Rudolph.
\newblock {\em Nature Phys.}, 8:475, 2012.

\bibitem{psi2}
P.~G. Lewis, D.~Jennings, J.~Barrett, and T.~Rudolph.
\newblock {\em Phys. Rev. Lett.}, 109:150404, 2012.

\bibitem{psi3}
R.~Colbeck and R.~Renner.
\newblock {\em Phys. Rev. Lett.}, 108:150402, 2012.

\bibitem{psi4}
L.~Hardy.
\newblock {\em Int. J. Mod. Phys. B}, 27:1345012, 2013.

\bibitem{psi5}
M.~K. Patra, S.~Pironio, and S.~Massar.
\newblock {\em Phys. Rev. Lett.}, 111:090402, 2013.

\bibitem{psi6}
M.~S. Leifer.
\newblock {\em Phys. Rev. Lett.}, 112:160404, 2014.

\bibitem{psi7}
J.~Barrett, E.~G. Cavalcanti, R.~Lal, and O.~J.~E. Maroney.
\newblock {\em Phys. Rev. Lett.}, 112:250403, 2014.

\bibitem{timpson2010}
David Wallace and Christopher~G. Timpson.
\newblock Quantum mechanics on spacetime i: Spacetime state realism.
\newblock {\em The British Journal for the Philosophy of Science},
  61(4):697--727, 2010.

\bibitem{butterfield1992}
J.~Butterfield.
\newblock Bell's theorem: What it takes.
\newblock {\em British Journal for the Philosophy of Science}, 43:41, 1992.

\bibitem{brillouin1952}
L.~Brillouin.
\newblock {\em A physical theory of information}.
\newblock Watson Laboratories, 1952.

\bibitem{deutsch2014}
D.~Deutsch and C.~Marletto.
\newblock Constructor theory of information.
\newblock {\em Proceedings of the Royal Society of London A: Mathematical,
  Physical and Engineering Sciences}, 471(2174), 2014.

\bibitem{montina2008}
A.~Montina.
\newblock Exponential complexity and ontological theories of quantum mechanics.
\newblock {\em Phys. Rev. A}, 77(2):022104, 2008.

\bibitem{timpson-qm}
C.~Timpson.
\newblock Philosophical aspects of quantum information theory.
\newblock In D.~Rickles, editor, {\em The Ashgate Companion to Contemporary
  Philosophy of Physics}, page 197. Ashgate, 2008.

\bibitem{bennett1992}
C.~H. Bennett and S.~J. Wiesner.
\newblock Communication via one- and two-particle operators on
  einstein-podolsky-rosen states.
\newblock {\em Phys. Rev. Let.}, 69(20):2881, 1992.

\bibitem{cerf1996}
N.~Cerf and C.~Adami.
\newblock Quantum information theory of entanglement and measurement.
\newblock {\em Physica D}, 120:62, 1998.

\end{thebibliography}

\end{document}